\documentclass[twocolumn,twoside,slac_two]{revtex4}
\usepackage{graphicx}
\usepackage{fancyhdr}
\usepackage{amsmath}
\input babarsym.tex

\newcommand{\hz}{\ensuremath{h^0}\xspace}
\newcommand{\sinbb}{\ensuremath{\sin2\beta}\xspace}
\newcommand{\cosbb}{\ensuremath{\cos2\beta}\xspace}
\newcommand{\sinbbeff}{\ensuremath{\sin2\beta_\mathrm{eff}}\xspace}

\newcommand{\betaeff}{\ensuremath{\beta_\mathrm{eff}}\xspace}
\newcommand{\abslambda}{\ensuremath{|\lambda|}\xspace}
\def\calC{{\ensuremath{\cal C}}\xspace}
\newcommand{\dm}{\ensuremath{\Delta m}\xspace}
\newcommand{\dt}{\ensuremath{\Delta t}\xspace}
\newcommand{\Btag}{\ensuremath{B_\mathrm{tag}}\xspace}

\def\etal{{\it et al.}}
\newcommand{\A}{\ensuremath{{\cal A}}\xspace}
\newcommand{\Abar}{\ensuremath{\overline{\A}\xspace}}

\newcommand{\Af}{\ensuremath{\A_f}\xspace}
\newcommand{\Abarf}{\ensuremath{\Abar_f}\xspace}
\newcommand{\lplm}{\ensuremath{\ell^+\ell^-}\xspace}

\newcommand{\AD}{\ensuremath{\A_D}\xspace}
\newcommand{\ADbar}{\ensuremath{\A_{\Db}}\xspace}

\newcommand{\msqKsp}{\ensuremath{m^{2}_{\KS\pip}}\xspace}

\newcommand{\msqKsm}{\ensuremath{m^{2}_{\KS\pim}}\xspace}
\newcommand{\etap}{\ensuremath{\eta^{\prime}}\xspace}

\begin{document}

\title{Measurements of the CKM Angle {\boldmath $\beta/\phi_1$} at {\boldmath
    $B$} Factories}

%

\author{C.~H.~Cheng for the \babar\ and Belle collaborations}
\affiliation{California Institute of Technology, Pasadena, California 91125, USA}

\begin{abstract}
We present a review of the measurements of the CKM angle $\beta$
$(\phi_1)$
by the \babar\ and Belle experiments at the asymmetric-energy \epem
$B$ Factories PEP-II and KEK$B$. The angle $\beta$ ($\phi_1$) is measured by
time-dependent \CP analyses of neutral \B meson decays in a $\FourS\ra\BB$
system, where one \B meson is fully reconstructed in a final state that can be
accessed to by both \Bz and \Bzb, usually a \CP eigenstate. This angle
has been measured at a high precision through $\Bz\ra (\ccbar)\Kz $ channels.
We also review another tree-dominated decay $\Bz\ra D^{(*)0}\hz$ ($\hz = \piz,\,
\eta^{(\prime)},\, \omega$); tree decays with penguin pollutions, $\Bz\ra
D^{(*)\pm}D^\mp$ and $J/\psi \piz$; and penguin dominated modes,
$\Bz\ra \eta^\prime\Kz$, $\Kp\Km\Kz$, and $\KS\KS\KS$. A hint of \sinbb ($\sin
2\phi_1$) in 
charmless modes less than $(\ccbar)\Kz$ modes still persists, which may be an
indication of possible new physics entering the loop in the penguin diagram.
\end{abstract}

\maketitle

\thispagestyle{fancy}


\section{Introduction}

Measurements of time-dependent \CP asymmetries in \Bz meson decays, through
the interference between decays with and without \Bz-\Bzb mixing, have
provided stringent tests on the mechanism of \CP violation in the standard
model (SM). The time-dependent \CP asymmetry amplitude equals to \sinbb
\footnote{\babar\ and Belle collaborations use different conventions
to label the three CKM angles. Here we adopt the \babar\ convention of $\beta$,
$\alpha$ and $\gamma$, rather than Belle's $\phi_1=\beta$, $\phi_2=\alpha$, 
and $\phi_3=\gamma$. } 
 in the
SM if the
\B meson decays to a final states that can be accessed to by both \Bz and \Bzb
without non-trivial relative weak phase. The angle $\beta =
\mathrm{arg}(V_{cd}V_{cb}^*/V_{td}V_{tb}^*)$ is a phase in the 
Cabibbo-Kobayashi-Maskawa (CKM) quark-mixing matrix~\cite{CKM}. The phase
difference, $2\beta$, between decays with and without \Bz-\Bzb mixing, arises
through the box diagrams in \Bz-\Bzb mixing, which are dominated by the
diagrams with virtual top quark.

In this paper we present a review of recent measurements of $\beta$ at \babar\
and Belle experiments at the asymmetric-energy \epem $B$ Factories PEP-II and
KEK$B$, including precision measurements using $\Bz\ra (\ccbar)\Kz$ decays,
new decay modes with tree diagram, $\Bz\ra D^{(*)0}\hz$ ($\hz = \piz,\,
\eta^{(\prime)},\, \omega$), tree decays with penguin pollutions, $\Bz\ra
D^{(*)\pm}D^\mp$ and $J/\psi \piz$, and penguin dominated modes,
$\Bz\ra \eta^\prime\Kz$, $\Kp\Km\Kz$, and $\KS\KS\KS$. We also present the
measurements that resolve the two-fold ambiguity in $2\beta$. Finally we
present the comparison between the latest $\sinbb$ measurements of
$(\ccbar)\Kz$ modes and charmless modes. 

At the time of the writing of this paper, Belle and \babar\ have collected
more than 700\invfb and 400\invfb of data, respectively, which correspond to
more than 1.1 billion $\FourS\ra\BB$ decays. Most results shown here are based on
$535\times 10^6$ ($383\times 10^6$) \BB pairs of data for Belle (\babar)
experiment. 

\section{Time-Dependent {\boldmath \CP} Asymmetry}

To measure time-dependent \CP asymmetries, we typically fully reconstruct a
neutral \B meson decaying into a \CP eigenstate. 
From the remaining particles in
the event, the vertex of the other \B meson, \Btag, is reconstructed and
its flavor is identified (tagged). The proper decay time difference $\dt=
t_{\CP}- t_\mathrm{tag}$, between the signal \B ($t_{\CP}$) and \Btag
($t_\mathrm{tag}$) is determined from the measured distance between the
two \B decay vertices projected onto the boost axis
and the boost ($\beta\gamma= 0.56$ at PEP-II and $0.43$ at KEK$B$) of the
center-of-mass (c.m.) system. The \dt distribution, assuming \CP conservation
in \Bz-\Bzb mixing and $\Delta\Gamma/\Gamma=0$, is
given by:
\begin{align}
F_\pm(\dt) & = \frac{\Gamma e^{-\Gamma|\dt|}}{4} [ 1\mp \Delta w \pm
 \label{eq:fdt}\\
 & (1-2w) (\eta_f \calS \sin(\dm\dt) - \calC \cos(\dm\dt)) ]\,,  \notag
\end{align}
where the upper
(lower) sign is for events with \Btag being identified as a \Bz (\Bzb),
 $\eta_f$ is the \CP eigenvalue of the final state,
\dm is the \Bz-\Bzb mixing frequency, $\Gamma$ is the mean decay rate of
the neutral \B meson, the mistag parameter $w$ is the probability of
incorrectly identifying the flavor of \Btag, and $\Delta w$ is the difference
of $w$ for \Bz and \Bzb. In the SM, the parameters $\calS =
\mathrm{Im}\lambda/(1+\abslambda)$ and $\calC =
(1-\abslambda)/(1+\abslambda)$, where $\lambda=
\frac{q}{p}\frac{\Abarf}{\Af}$, and \Af (\Abarf) is the amplitude of \Bz
(\Bzb) decaying to the \CP final state $f$. In the SM, if only one diagram
contributes to the decay process, $\calS= -\sinbb$ and $\calC =0$. A non-zero
value of \calC would indicate direct \CP violation.

Because there can be other diagrams with a different weak phase, the
experimental result of \calS does not necessarily equal to $-\sinbb$. To
separate the measured value from  the standard model \sinbb, we denote the
measured one \sinbbeff.

\section{\boldmath $\Bz\ra(\ccbar)\Kz $}

The \CP violation in neutral \B meson system was first established
experimentally by 
\babar\ and Belle in 2001 using \B decays to a charmonium $(\ccbar)$ and a
neutral $K$ meson~\cite{firstCP}. These modes are dominated by a
color-suppressed $\b\ra\ccbar\s$ tree diagram. The dominant penguin diagram
has the same weak phase. The term that has a different weak phase is a penguin
contribution that is Cabibbo
suppressed by ${\cal O}(\sin^2\theta_{\mathrm{Cabibbo}})$. Therefore, to a
good accuracy, the parameters $\calS = -\sinbb$ and $\calC=0$. Recent
theoretical calculations suggest that the correction on \calS is in the order of
$10^{-3}$--$10^{-4}$~\cite{cck0corr}. 

These modes have relatively large (${\cal O}(10^{-4}$--$10^{-5})$). 
branching fractions and have low experimental backgrounds and high
reconstruction efficiencies. For the mode 
$\Bz\ra J/\psi(\lplm)\KS(\pip\pim)$, the signal purity is typically
greater than 95\%. 

From $535\times 10^{6}$ \BB pairs, Belle reconstructs approximately 7500
$\Bz\ra J/\psi\KS$ and 6500 $J/\psi \KL$ signal events, with $J/\psi\ra\lplm$
($\ell= e\,,\mu$) and $\KS\ra\pip\pim$ or $\piz\piz$, and measures $\sinbbeff=
+0.642\pm 0.031\pm 0.017$ and $\calC = -0.018\pm 0.021\pm
0.014$~\cite{Chen:2006nk}. 

\babar\ uses many other modes in addition to $ J/\psi\KS$ and $J/\psi\KL$, including
$\psi(2S)\KS$, $\chi_{c1}\KS$, $\eta_c\KS$ and $J/\psi
\Kstarz(\KS\piz)$. \babar\ reconstructs approximately 6900 \CP-odd signal events and
3700 \CP-even signal events from $383\times 10^{6}$ \BB pairs,
and obtains $\sinbbeff=
+0.714\pm 0.032\pm 0.018$ and $\calC = +0.049\pm 0.022\pm
0.017$~\cite{Aubert:2007hm}.
In addition, \babar\ also performs measurements, including
systematic uncertainties, using individual mode, because the theoretical
corrections   could in
principle be different among those modes. \babar\ also reports the result
using the same decay modes as used by Belle in order to provide a direct
comparison. The two experiments agree well within the uncertainty. 

The averages, calculated by the Heavy Flavor Averaging Group ({\ttfamily
  HFAG})~\cite{HFAG}, are  $\sinbbeff= +0.678\pm 0.026$ and $\calC = +0.012\pm
0.020$. The 
uncertainty on \sinbbeff is 4\%, or approximately $1^\circ$ on $\beta$. 
Because of the high experimental precision and low theoretical uncertainty,
the result from these modes serves as a benchmark in the SM; any other
measurements of \sinbb that have a significant deviation from it, beyond the
usually small SM corrections, would indicate evidence for new physics.

\section{\boldmath $\Bz\ra D^{(*)\pm}D^{(*)\mp} $}

The decay $\Bz\ra D^{(*)\pm}D^{(*)\mp} $ is dominated by a color-allowed
Cabibbo-suppressed $\b\ra \ccbar \d$ tree diagram. The penguin diagram in the
SM has a different weak phase and is expected to contribute few percent
correction~\cite{xing} to \CP asymmetry. A large deviation in \sinbbeff
from that in $\Bz\ra(\ccbar)\Kz$ would indicate possible new physics
contribution to the loop in the penguin diagram.

The final state $\Dp\Dm$ is a \CP eigenstate so $\calS= -\sinbb$ and $\calC=0$
in the SM when neglecting the penguin contribution. The final state
$D^{*\pm}D^{*\mp}$ is a mixture of \CP even and \CP odd states. An angular
analysis is needed to disentangle the contributions from different \CP
states. The final state $D^{*\pm}D^{\mp}$ is not a \CP eigenstate. The decay
amplitudes can have a strong phase difference $\delta$, i.e., 
$\A(\Bz\ra \Dstarp\Dm)/\A(\Bz\ra \Dstarm\Dp) = R e^{i\delta}$. As a result,
the \calS and \calC parameters, ($+$ for $\Dstarp\Dm$ and 
$-$ for $\Dstarm\Dp$) are 
$\calS_\pm = 2 R \sin(2\beta_{\mathrm{eff}}\pm\delta)/(1+R^2)$, and 
$\calC_\pm = \pm (R^2-1)/(R^2+1)$, assuming there is no direct \CP violation. 

\begin{figure}[h]
\centering
\includegraphics[width=37mm]{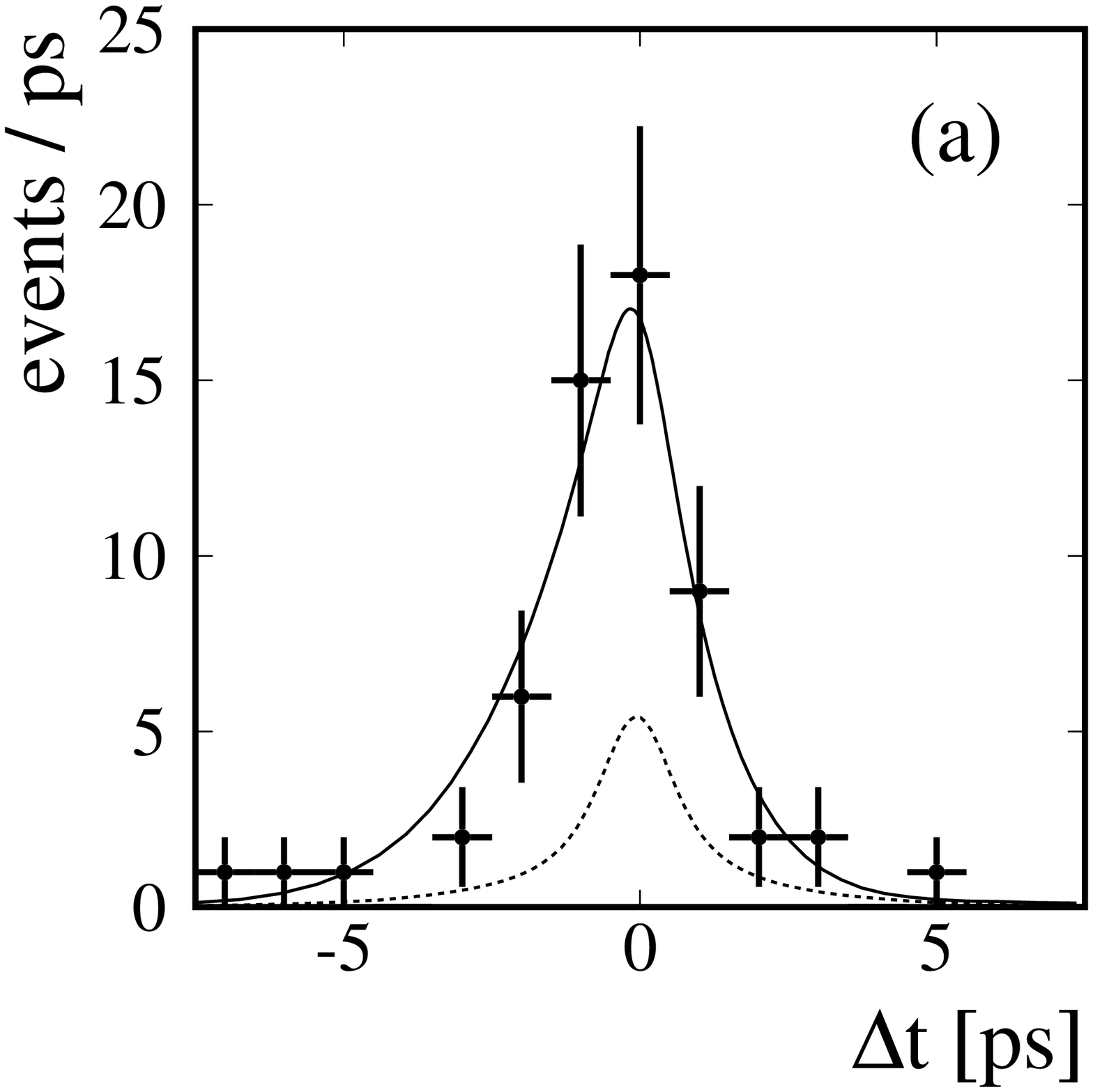}
\includegraphics[width=37mm]{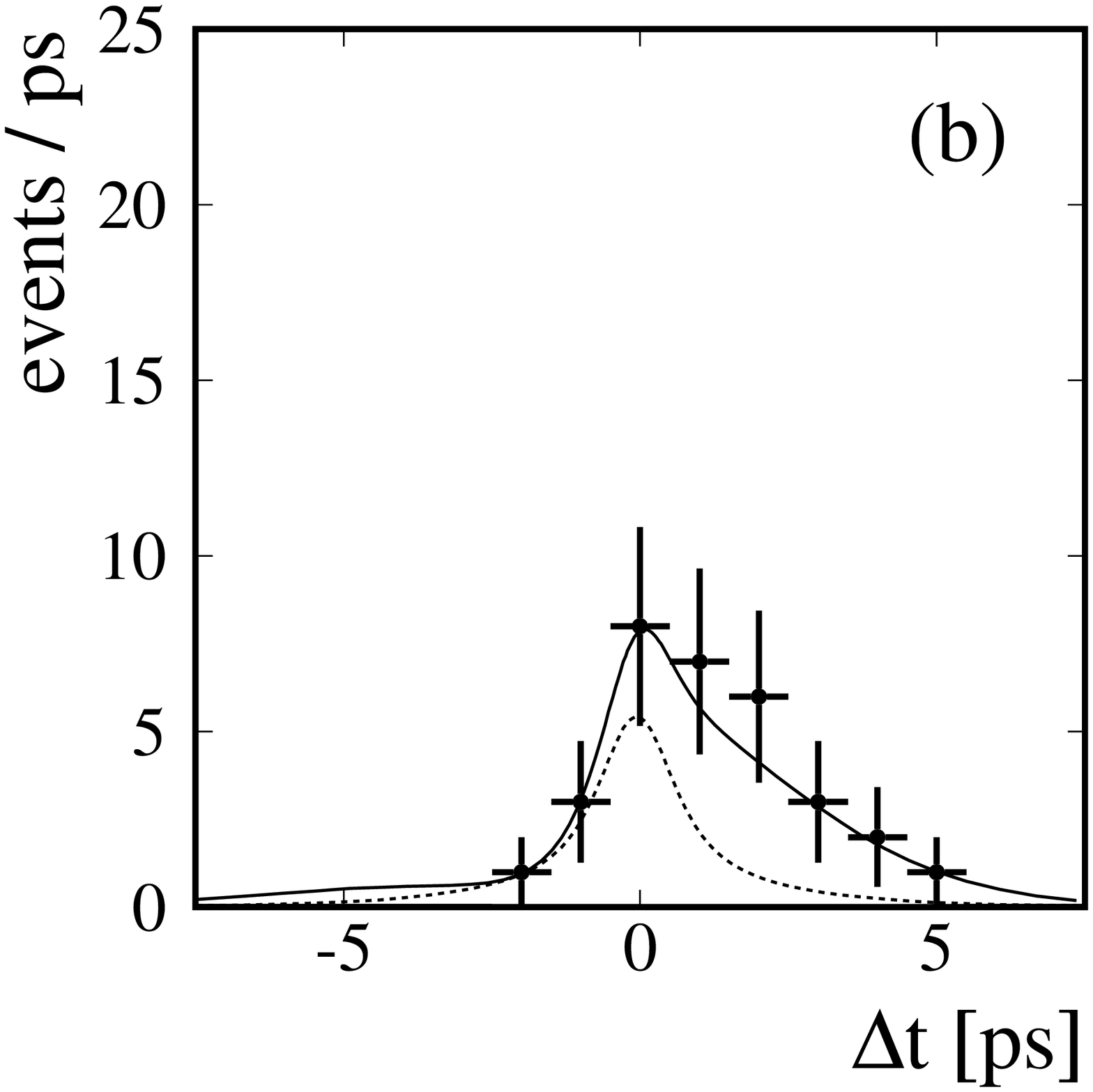}
\includegraphics[width=80mm]{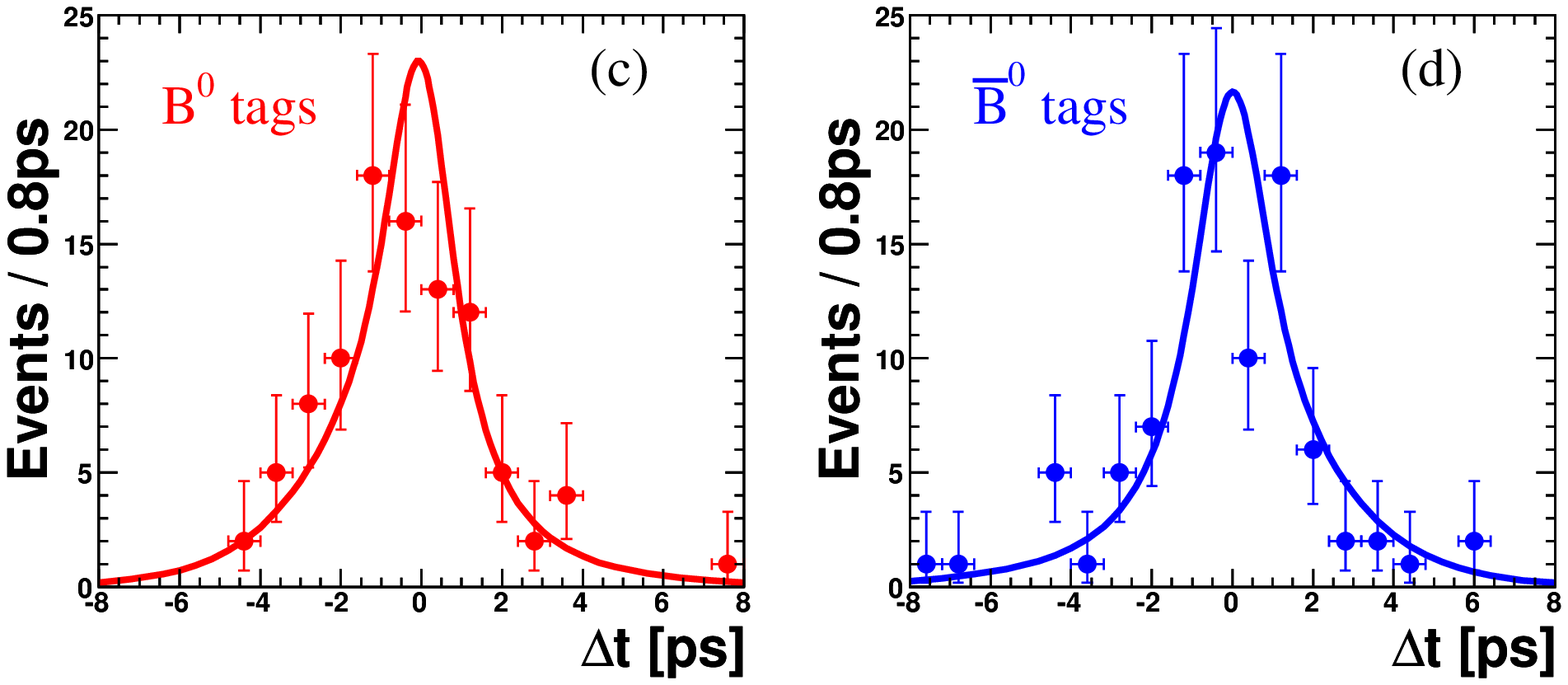}
\caption{\dt distributions of $\Bz\ra\Dp\Dm$ from (a,b) Belle experiment and
  (c,d) \babar\ experiment. Plots (a) and (c) are for \Bz-tagged events and
  (b) and (d) are for \Bzb-tagged events. 
\label{fig:DpDm}}
\end{figure}

\begin{figure}[h]
\centering
\includegraphics[width=80mm]{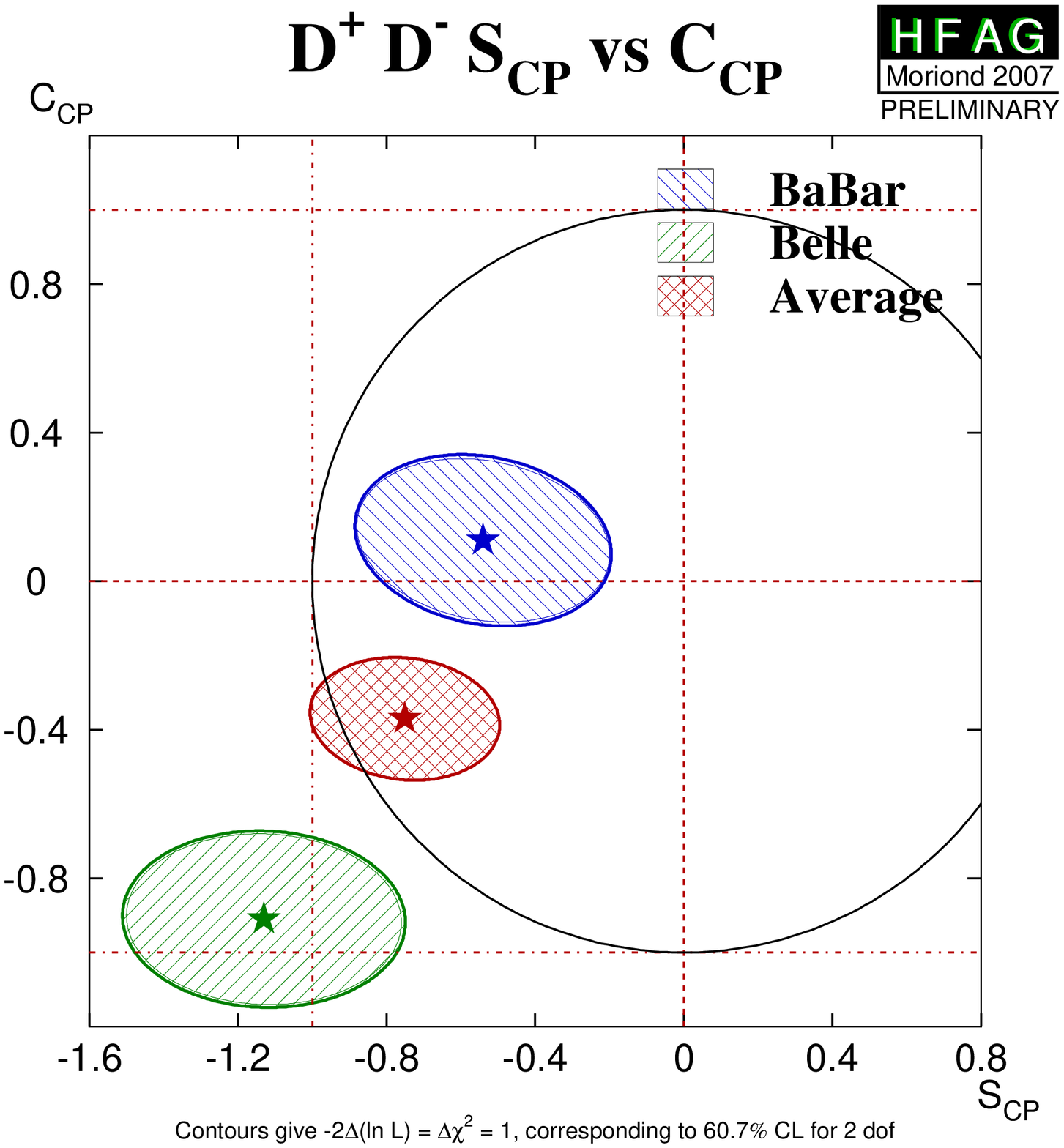}
\caption{The comparison between Belle's and \babar's
  $\Bz\ra\Dp\Dm$ results. 
\label{fig:DpDm-SC}}
\end{figure}

Belle collaboration recently reports evidence for a large direct \CP violation
in $\Dp\Dm$ channel using a data sample of $535\times 10^{6}$ \BB pairs. They
measure $\calS= -1.13\pm 0.37\pm 0.09$ and $\calC=
-0.91\pm0.23\pm0.06$~\cite{BelleDpDm}. The \CP conservation, $\calS=\calC=0$,
is excluded at $4.1\sigma$ level and $\calC=0$ is excluded at $3.2\sigma$.
However, such a large direct \CP violation has not been observed in previous
measurements with  $\Bz\ra D^{(*)\pm}D^{(*)\mp} $ decays, involving the same
quark-level weak
decay~\cite{Aubert:2005rn,Aubert:2005hs,Aushev:2004uc,Miyake:2005qb}. 
\babar\ also uses the same decay modes and measures 
$\calS = -0.54\pm0.34\pm0.06$ and $\calC=
0.11\pm0.22\pm0.07$~\cite{BabarDpDm}, which is 
consistent with the SM with small penguin contributions. 
Figure~\ref{fig:DpDm} shows the \dt distributions for \Bz-tagged and
\Bzb-tagged events separately. For the result from Belle, the plots show clear
difference between the yields of \Bz-tagged and \Bzb-tagged events. The
consistency of these two results are quite low ($\chi^2/\mathrm{dof}= 12/2$,
or C.L.=0.003, corresponding to $3\sigma$). Figure~\ref{fig:DpDm-SC} shows the
comparison by {\ttfamily HFAG}. 

\babar\ also reports an updated measurement of $\Bz\ra \Dstarpm
\Dmp$~\cite{BabarDpDm}. The result is
$\calC_{\Dstarp\Dm} = 0.18\pm 0.15 \pm 0.04$, 
$\calS_{\Dstarp\Dm} = -0.79\pm 0.21 \pm 0.06$,
$\calC_{\Dstarm\Dp} = 0.23\pm 0.15 \pm 0.04$, and
$\calS_{\Dstarm\Dp} = -0.44\pm 0.22 \pm 0.06$. 
It is
also illustrative to express the parameters \calC
and \calS in a slightly different parameterization~\cite{Aubert:2003wr}:
$\calC_{\Dstar D}=(\calC_{\Dstarp\Dm}+\calC_{\Dstarm\Dp})/2$, $\Delta \calC_{\Dstar
D}=(\calC_{\Dstarp\Dm}-\calC_{\Dstarm\Dp})/2$, $\calS_{\Dstar
D}=(\calS_{\Dstarp\Dm}+\calS_{\Dstarm\Dp})/2$ and $\Delta \calS_{\Dstar
D}=(\calS_{\Dstarp\Dm}-\calS_{\Dstarm\Dp})/2$.  The quantities $\calC_{\Dstar D}$
and $\calS_{\Dstar D}$ parameterize flavor-dependent direct \CP violation,
and mixing-induced \CP violation related to the angle $\beta$,
respectively.  The parameters $\Delta \calC_{\Dstar D}$ and $\Delta
\calS_{\Dstar D}$ are insensitive to \CP violation. $\Delta \calC_{\Dstar D}$
describes the asymmetry between the rates
$\Gamma(\Bz\to\Dstarp\Dm)+\Gamma(\Bzb\to\Dstarm\Dp)$ and
$\Gamma(\Bz\to\Dstarm\Dp)+\Gamma(\Bzb\to\Dstarp\Dm)$, while $\Delta
\calS_{\Dstar D}$ is related to the strong phase difference, $\delta$. \babar\ 
reports
\begin{align}
\calC_{\Dstar D} &= \;\;\;0.21\pm 0.11 \pm 0.03  \nonumber \\
\calS_{\Dstar D} &= -0.62\pm 0.15 \pm 0.04  \nonumber \\
\Delta \calC_{\Dstar D} &= -0.02\pm 0.11 \pm 0.03  \nonumber \\
\Delta \calS_{\Dstar D} &= -0.17\pm 0.15 \pm 0.04  \,.
\end{align}
The parameter $\calS_{\Dstar D} = \frac{2R}{1+R^2}\cos\delta \sinbbeff$ in the
SM. \babar\ finds that it is non-zero at approximately $4\sigma$ level, which
indicates $ \sinbbeff\neq 0$ at the same significance in these modes.

\section{\boldmath $\Bz\ra J/\psi \piz$}

The $\Bz\ra J/\psi \piz$ decay has the same quark level diagrams as $J/\psi\Kz$
except that the \s quark in $\b\ra\ccbar\s$ is substituted by a \d
quark. Therefore, the dominant tree diagram is Cabibbo suppressed compared to
that of $J/\psi\Kz$. However, unlike $J/\psi\Kz$, the dominant penguin diagram
in $J/\psi\piz$, whose CKM element factor is in the same order as the tree,
has a different weak phase from the tree. Therefore the deviation in \sinbbeff
from that of $J/\psi\Kz$ could be substantial.
This mode is also useful to constrain the penguin pollution in
$\Bz\ra(\ccbar)\Kz$ mode in a more model-independent way (assuming
 SU(3) symmetry)~\cite{Ciuchini:2005mg}. 

Neither \babar\ nor Belle updated their results since 2005. The current
results are $\sinbbeff= -0.72\pm0.42\pm0.09$ and $\calC = 0.01\pm0.29\pm0.03$
using $152\times 10^{6}$ \BB pairs in Belle~\cite{Kataoka:2004mt} and
$\sinbbeff= -0.68\pm0.30\pm0.04$ and $\calC= -0.21\pm0.26\pm0.06$ using  
$232\times 10^{6}$ \BB pairs in \babar~\cite{Aubert:2006qc}.

\section{\boldmath $\Bz\ra D^{(*)0}\hz$ ($\hz = \piz,\,\eta^{(\prime)},\,\omega$)}

\subsection{\boldmath \sinbbeff measurement using \Dz decays to \CP
  eigenstates} 

\begin{figure}[htb]
\begin{center}
\includegraphics[width=50mm]{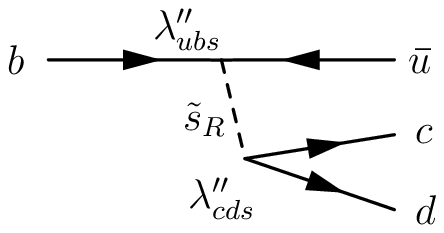}
\caption{$R$-parity-violating supersymmetry diagram that could
  contribute to $\Bz\ra D^{(*)0}\hz$ decays.}
\label{fig:Rpsusy}
\end{center}
\end{figure}

The decay $\Bz\ra D^{(*)0}\hz$ ($\hz = \piz,\,\eta^{(\prime)},\,\omega$) is
dominated by a color-suppressed $\b\ra\c\ubar\d$ tree diagram. The final state
is a \CP eigenstate if the neutral
$D$ meson also decays to a \CP eigenstate, and therefore Eq.~\ref{eq:fdt}
applies. This mode is free of penguin diagrams. The next diagram is also a
color-suppressed tree diagram, $\b\ra\u\cbar\d $, which is doubly Cabibbo
suppressed. The SM correction on \sinbbeff is believed to be a few
percent~\cite{Grossman:1996ke}. Because it has no penguin contributions, the
``usual'' new physics that only enters the loops through unobserved heavy
virtual particles would not affect these decays, other than that they can
still affect 
the box diagrams in \Bz-\Bzb mixing. However, more exotic new physics models
such as $R$-parity-violating supersymmetry~\cite{Grossman:1996ke} could enter at
tree level in these decays (Fig.~\ref{fig:Rpsusy}).

\babar\ recently reports a measurement of \sinbbeff using $D^{(*)0}\piz$ with
$\Dz\ra\Kp\Km,\,\KS\omega$, and $D^{(*)0}\eta$ with $\Dz\ra\Kp\Km$, and
$\Dz\omega$ with $\Dz\ra\KpKm,\,\KS\omega,\,\KS\piz$. The \Dstarz is
reconstructed from $\Dstarz\ra\Dz\piz$, when applicable. \babar\ uses
$383\times 10^6$ 
\BB pairs and obtains $\sinbbeff = 0.56\pm 0.23\pm 0.05$ and $\calC =
-0.23\pm0.16\pm0.04$~\cite{Aubert:2007mn}. This result is $2.3\sigma$ away
from \CP conservation.

\subsection{Resolve ambiguity using {\boldmath $\Dz\ra\KS\pip\pim $}}

The \Bz mixing phase $2\beta$ has a two-fold ambiguity from \sinbb measurement,
$2\beta$ and $\pi-2\beta$ 
(or equivalently $\beta$ has a four-fold ambiguity, $\beta$, $\pi/2-\beta$,
$\pi+\beta$, and $3\pi/2-\beta$). The ambiguity can be resolved by studying
decay modes that involve multi-body final states, where the known variation of
the strong phase differences across the phase space allows one to access
\cosbb in addition to \sinbb. To resolve the ambiguity, one only need to know
the sign of \cosbb. 

Both Belle and \babar\ have performed a time-dependent $D\ra\KS\pip\pim$
Dalitz plot analysis in the decay $\Bz\ra
D^{(*)}[\KS\pip\pim]\hz$~\cite{Bondar:2005gk} to measure \cosbb (and
\sinbb). The decay rate of the \B meson, accompanied by a \Bz ($+$ sign) or
\Bzb ($-$ sign) is proportional to 
\begin{align}
\label{eq:kpp}
{\cal P}_\pm & = \frac{e^{-\Gamma\dt}}{2} |\A_B|^2 \cdot
\Big[ (|\ADbar|^2+ |\lambda|^2 |\AD|^2) \notag \\
& \mp  (|\ADbar|^2- |\lambda|^2 |\AD|^2) \cos(\dm\dt)  \\
& \pm 2|\lambda| \xi_{\hz} (-1)^L
\mathrm{Im}(e^{-2i\beta}\AD\ADbar^{*})\sin(\dm\dt)\Big]\,,  \notag
\end{align}
where $\A_B$ is the \B decay amplitude, and $\AD$ (\ADbar) is the decay
amplitude of \Dz (\Dzb) and is a function of the Dalitz plot
variables $(\msqKsp,\msqKsm)$, which is determined from large data samples of
$\epem\ra X \Dstarp$, $\Dstarp\ra \Dz \pip$ events. 
The factor $\xi_{\hz}$ is the \CP eigenvalue of \hz, and $(-1)^L$ is the
angular momentum factor.
In the last term of Eq.~\ref{eq:kpp} we can rewrite
\begin{align}
\mathrm{Im}(e^{-2i\beta}\AD\ADbar^{*}) & = \mathrm{Im}(\AD\ADbar^*)\cosbb
\notag \\
& - \mathrm{Re}(\AD\ADbar^*)\sinbb\,,
\end{align}
and treat \cosbb and \sinbb as independent parameters in the analyses.

Belle obtains $\cosbb= 1.87^{+0.40+0.22}_{-0.53-0.32}$ and
$\sinbb=0.78\pm0.44\pm0.22$, and determines $\cosbb>0$ at 98.3\% confidence
level~\cite{Krokovny:2006sv}. \babar\ measures $\cosbb= 0.54\pm 0.54\pm
0.08\pm 0.18$ and $\sinbb= 0.45\pm0.36\pm0.05\pm0.07$, where the last errors
 are due to Dalitz model uncertainty, and 
determines $\cosbb>0$ at 87\% confidence~\cite{Aubert:2006an}.
Another mode ($\Bz\ra\Kp\Km\Kz$) can also be used to resolve this
ambiguity. We will discuss it later in Sec.~\ref{sec:kpkmk0}.

\section{\boldmath $\sinbbeff$ in $\b\ra\s$ penguin dominated modes}

In the measurement of \sinbb, different charmless modes have different
standard model corrections and uncertainties coming from, e.g.,
Cabibbo-suppressed trees, final state interaction long distance effect,
etc. Several theoretical calculations predict the corrections and
uncertainties are in the order of 1 to 10
percent~\cite{Beneke:2005pu,Williamson:2006hb,Cheng:2005bg}.

These charmless $\b\ra\s\qqbar$ penguin modes are more sensitive to new
physics that 
enters the loops because the new physics does not have to compete with the SM
tree processes. In this section we present several notable \sinbb measurements
in charmless \B decays and compare the current results with the high
precision $\B\ra(\ccbar)\Kz$ mode.

\subsection{\boldmath \Bz\ra\etap\Kz}

\begin{figure}[htb]
\centering
\includegraphics[height=40mm]{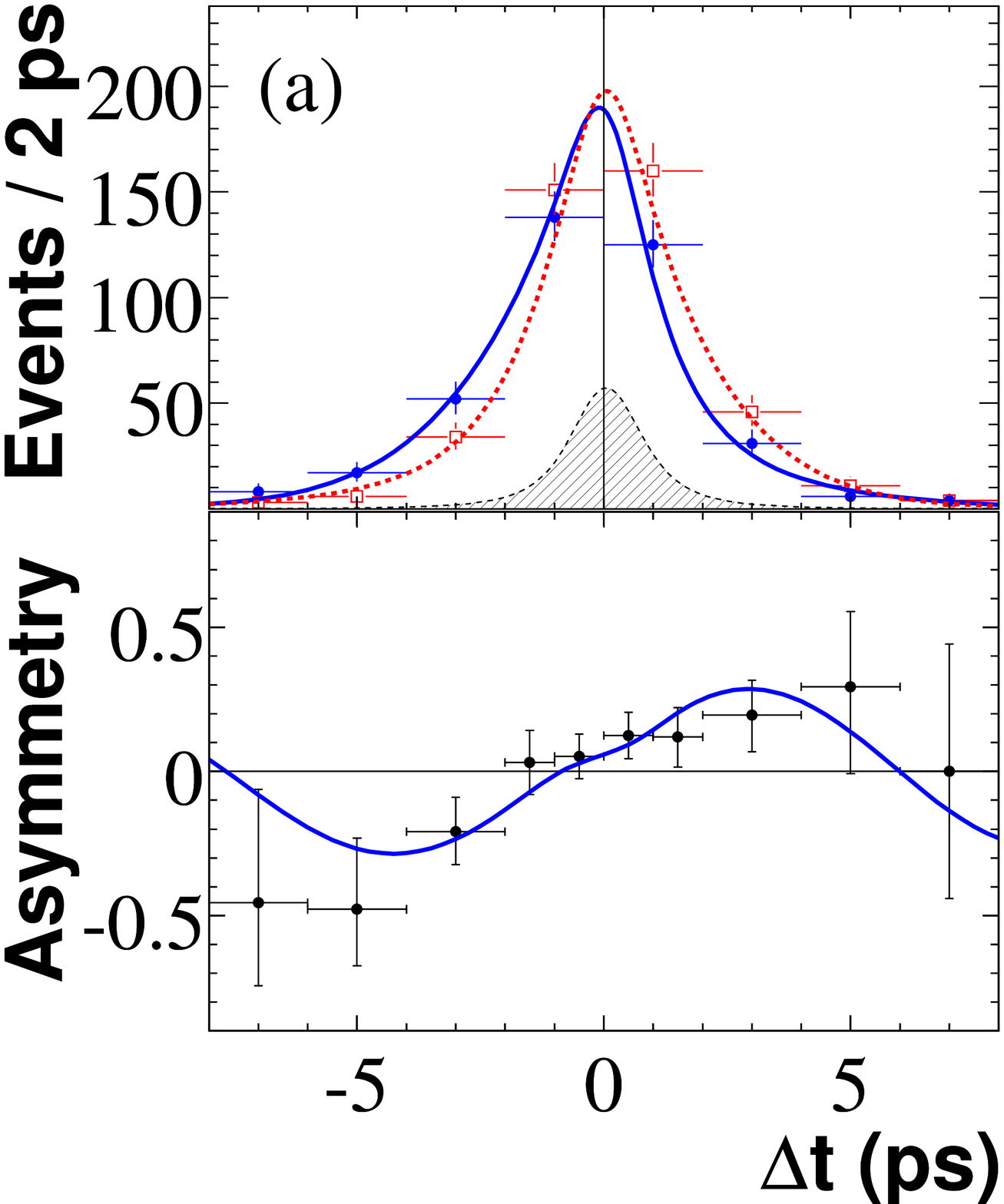}
\includegraphics[height=40mm]{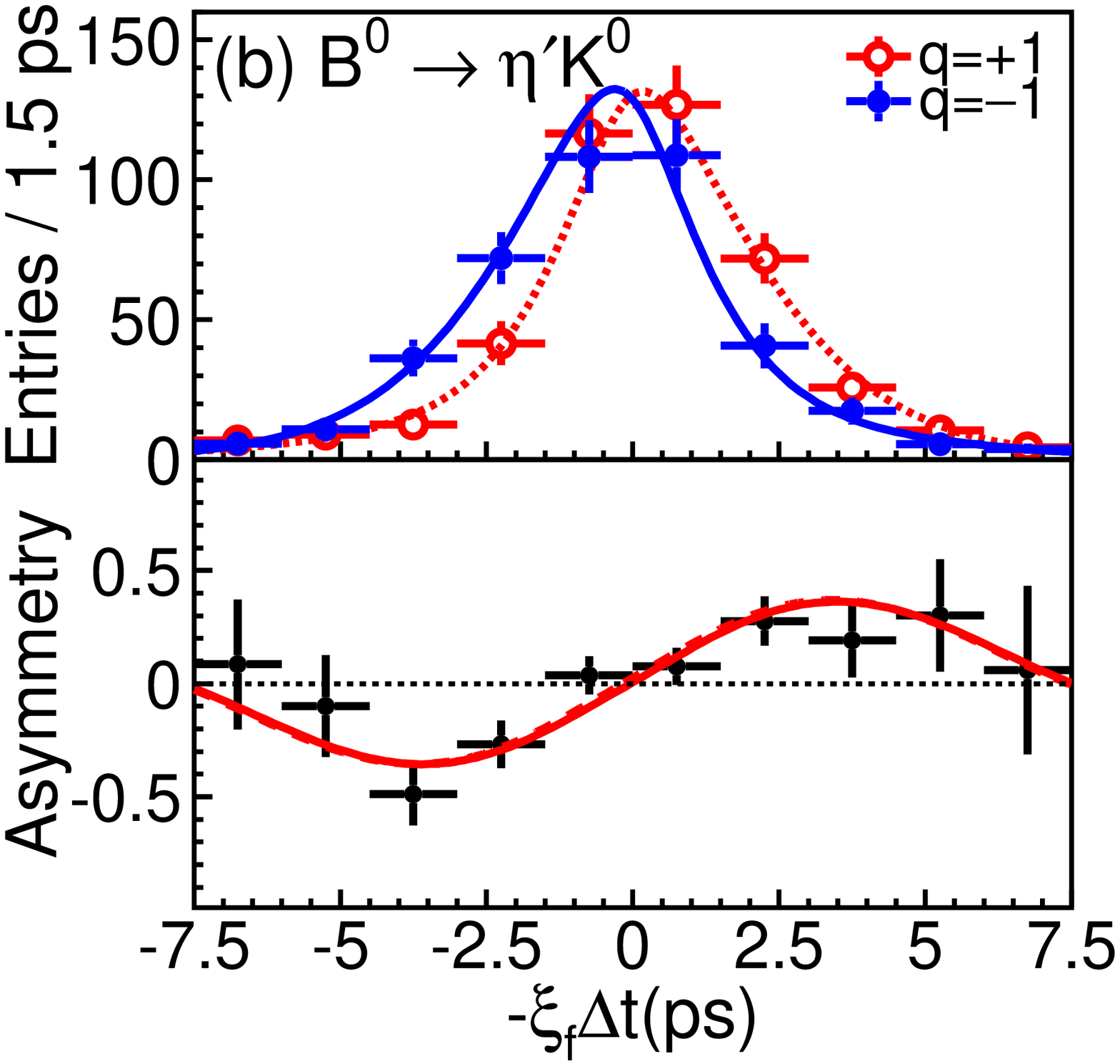}
\caption{The \dt distributions and asymmetries in $\etap\Kz$ mode for (a)
  \babar\ (only  $\etap\KS$ is shown) and (b) Belle.  
\label{fig:etaK0}}
\end{figure}

This mode is the most precisely measured penguin mode in the \B Factories. It
also has one of the smallest theoretical corrections and
uncertainties. Therefore it is arguably the best penguin mode for searches of
new physics that could affect \sinbb. Both \babar\ and Belle published their
observations of \CP asymmetry in this mode this year with more than $5\sigma$
significance. This is the first time \CP violation is observed in penguin
modes with such a large significance. \babar\ uses $383\times 10^6$ \BB pairs
($\sim1050$ $\etap\KS$ and $\sim250$ $\etap\KL$ signal events)
and measure $\sinbbeff= 0.58\pm 0.10\pm 0.03$ and $\calC= -0.16\pm 0.07\pm
0.03$~\cite{Aubert:2006wv}. Belle uses $535\times 10^6$ \BB pairs
($\sim1420$ $\etap\KS$ and $\sim450$ $\etap\KL$ signal events)
and measure $\sinbbeff= 0.64\pm 0.10\pm 0.04$ and $\calC= 0.01\pm 0.07\pm
0.05$~\cite{Chen:2006nk}. The \dt distributions and asymmetries are shown in
Fig.~\ref{fig:etaK0}.

\subsection{\boldmath $\Bz\ra\Kp\Km\Kz$ and $\phi\Kz$} 
\label{sec:kpkmk0}

The total
branching fraction of the three-body $\Bz\ra\Kp\Km\Kz$ decay  is relatively
large, 
about six times the dominant resonance  
$\phi(\ra\Kp\Km)\Kz$. The final state is a mixture of \CP even and \CP odd
states, depending on the spins of the resonances. \babar~\cite{Aubert:2006av} reconstructs about
1000 $\Kp\Km\KS$ and 500 $\Kp\Km\KL$ signal events from $347\times
10^6$ \BB pairs, and performs a time-dependent Dalitz plot analysis to extract
the \CP asymmetry parameter and Dalitz plot model parameters
simultaneously. The probability function of the \B meson decay rate is
similar to Eq.~\ref{eq:kpp}. Here \babar\ chooses to extract $\beta$ directly,
rather than to fit for \cosbb and \sinbb. \babar's Dalitz model includes
$\phi(1020)$, $f_0(980)$, $X_0(1500)$~\cite{x01500}, $\chi_{c0}$, and
non-resonance terms. Using the whole phase space, \babar\ obtains
$\betaeff= 0.361\pm0.079\pm 0.037$, which corresponds to 
$\sinbbeff= 0.66\pm 0.12\pm0.05$. \babar\ also fits for the region with
$m_{\Kp\Km} < 1.1\gevcc$ and obtains 
$\betaeff(\phi\Kz) = 0.06\pm 0.16\pm0.05$, and 
$\betaeff(f_0\Kz) = 0.18\pm 0.19\pm0.04$, which correspond to 
$\sinbbeff(\phi\Kz) = 0.12 \pm 0.31\pm 0.10$ and
$\sinbbeff(f_0\Kz) = 0.35 \pm 0.35\pm 0.08$. 

Since the angle \betaeff is extracted directly, the ($2\beta$,
$\pi-2\beta$) ambiguity can be resolved. \babar\ scans the angle \betaeff and
finds the change in the log likelihood from the best solution (near
$21^\circ$) to the second 
solution (near $69^\circ$) corresponds to $4.5\sigma$ statistical
significance, and thus rules out the second solution.

Belle does not perform the time-dependent Dalitz analysis. It uses a
quasi-two-body approach for $\phi\Kz$, and combines all resonances in
$\Kp\Km\Kz$, excluding $\phi\Kz$. Using $535\times 10^6$ \BB pairs, Belle
obtains 420 $\phi\Kz$ signal events, and $\sinbbeff= 0.50\pm 0.21\pm 0.06$ 
and $\calC= -0.07\pm 0.15 \pm 0.05$~\cite{Chen:2006nk}. 
For $\Bz\ra\Kp\Km\Kz$, Belle reconstructs 840 signal events and measures 
$\sinbbeff= 0.68\pm 0.15\pm 0.03^{+0.21}_{-0.13}$ and $\calC = 0.09\pm 0.10\pm
0.05$~\cite{Abe:2006gy}. 
The last uncertainty in \sinbbeff is due to the limited knowledge of
the \CP content in $\Bz\ra\Kp\Km\Kz$, which are determined from the branching
fractions of $\Bp\ra\Kp\KS\KS$ and $\Bz\ra\Kp\Km\Kz$ using isospin relations. 
This uncertainty does not exist in \babar's measurement because of the full
Dalitz analysis.

\subsection{\boldmath $\Bz\ra\KS\KS\KS$}

The decay $\Bz\ra\KS\KS\KS$ is a pure \CP-even state~\cite{ksksks}, therefore
a Dalitz analysis is not necessary. Belle uses $535\times 10^6$ \BB pairs and
obtains $\sinbbeff = 0.30\pm 0.32\pm 0.08$ and $\calC = -0.31\pm 0.20\pm
0.07$~\cite{Chen:2006nk}. \babar\ uses $383\times 10^6$ \BB pairs and
determines $\sinbbeff = 0.71\pm 0.24\pm 0.04$ and $\calC = 0.02\pm 0.21\pm
0.05$~\cite{Aubert:2007me}.

\section{Conclusion}

The measurement of \sinbb is a rich program at the $B$ Factories. 
A total of more than 900 million $\FourS\ra\BB$ pairs have been analyzed and
the $B$ Factories have achieved a precision of 4\% in \sinbb measurement,
$\sinbb = 0.678\pm0.026$, using
$\Bz\ra(\ccbar)\Kz$ decays. By studying the time-dependent evolution in
multibody final states, such as 
$\Bz\ra\Dz[\KS\pip\pim]\hz$ and $\Bz\ra\Kp\Km\Kz$
(and $\Bz\ra J/\psi \Kstarz(\KS\piz)$~\cite{Aubert:2004cp,Itoh:2005ks}, not
discussed here), 
the ambiguity in $\beta$ is 
resolved and we are confident that $\beta= (21.3\pm1.0)^\circ$ (in
[$0,\pi$]), rather than $(68.7\pm1.0)^\circ$. 

Belle observes evidence for large direct \CP asymmetry in $\Bz\ra\Dp\Dm$
channel. However, it is not confirmed by \babar, and none of the other
$D^{(*\pm)}D^{(*\mp)}$ modes, which have the same quark-level weak decays, show
large direct \CP violation. More data are needed to resolve this discrepancy.

\begin{figure}[htb]
\centering
\includegraphics[width=80mm]{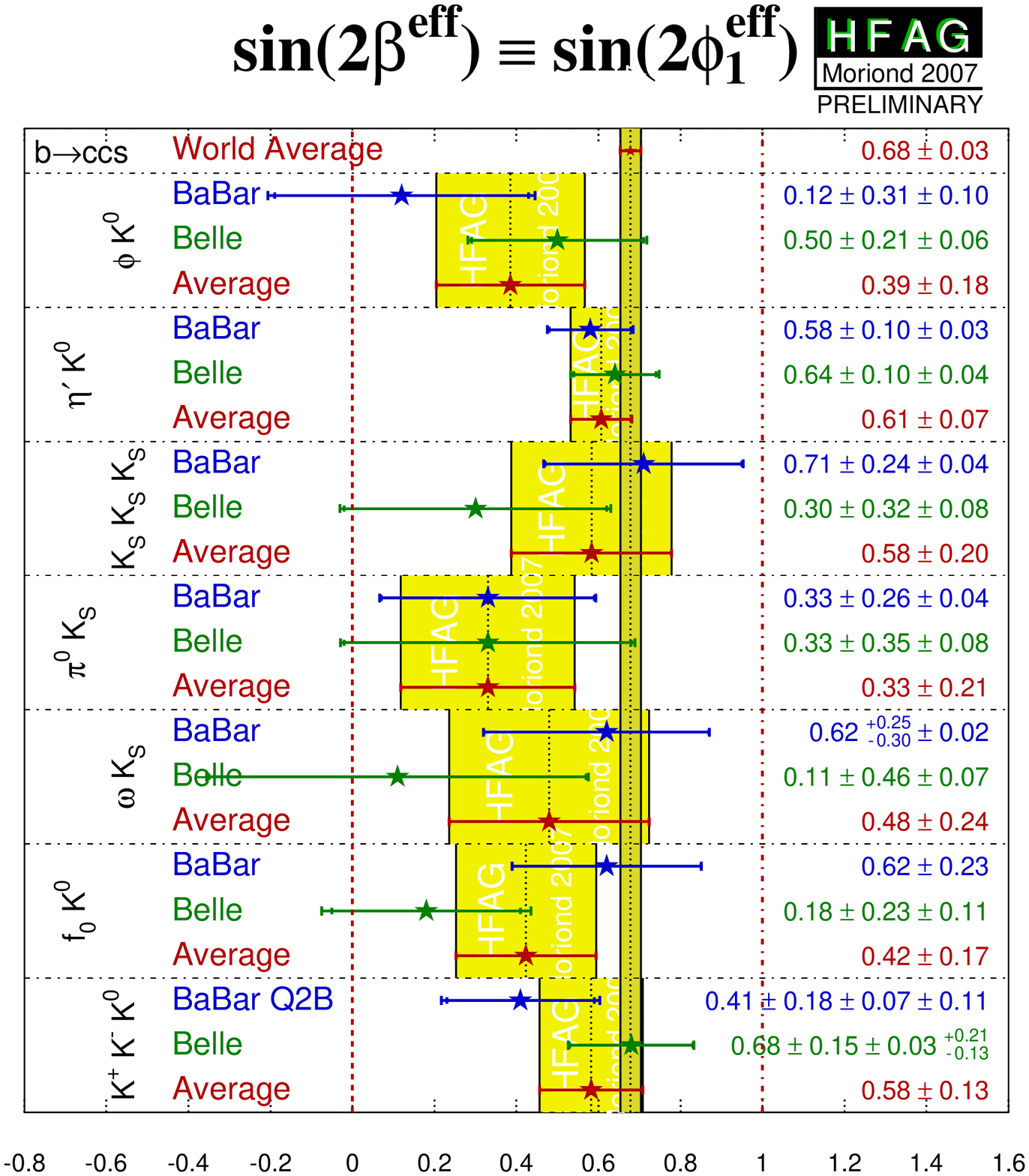}
\caption{Results of \sinbbeff from selected $\b\ra\s$ penguin modes, compared to
the world average, dominated by  $\Bz\ra(\ccbar)\Kz$ mode. \label{scp_pengWA}}
\end{figure}

Penguin dominated modes are great channels for probing physics beyond the
standard model. Both $B$ Factories have observed a clear \CP asymmetry in
$\Bz\ra\etap\Kz$ decays, and a great progress has been made in many other
penguin modes. 
Currently the central values of \sinbbeff in most of the penguin
modes are smaller than that in $(\ccbar)\Kz$ mode. See
Fig.~\ref{scp_pengWA}. The naive average of penguin modes, ignoring the
theoretical corrections and uncertainties, and the experimental correlations
among systematic uncertainties, is approximately 2.5 standard
deviations away from $(\ccbar)\Kz$ mode. This is a tantalizing hint of
possible new physics effect. However, it should not be taken too
seriously because of the aforementioned details we have ignored. 

Both $B$ Factories are expected to record more than double the analyzed
datasets before they finish the data taking in the next year or two. 
We will be able to constrain the standard model and physics beyond the
standard model much better, but it is unlikely we will have a
clear answer to whether new physics has a significant effect in the \CP
asymmetry in the \B decays.

\bigskip 

\end{document}